\documentclass[pra,twocolumn,floatix,amssymb,showpacs,amsmath,superscriptaddress]{revtex4-1}
\usepackage{graphicx}
\usepackage[usenames,dvipsnames]{color}
\usepackage{epsfig,color}
\usepackage{amsmath,amssymb}
\usepackage{mathptmx}
\usepackage[T1]{fontenc}
\usepackage{verbatim}
\usepackage{float}
\usepackage{units}
\usepackage{amsmath}
\usepackage{graphicx}
\usepackage[colorlinks = true,
            linkcolor = NavyBlue,
            urlcolor  = NavyBlue,
            citecolor = NavyBlue,
            anchorcolor = blue]{hyperref}
\usepackage{epstopdf}
\usepackage[normalem]{ulem}

\makeatletter

 \@ifundefined{textcolor}{}
 {
   \definecolor{BLACK}{gray}{0}
   \definecolor{WHITE}{gray}{1}
   \definecolor{RED}{rgb}{1,0,0}
   \definecolor{GREEN}{rgb}{0,1,0}
   \definecolor{BLUE}{rgb}{0,0,1}
   \definecolor{CYAN}{cmyk}{1,0,0,0}
   \definecolor{MAGENTA}{cmyk}{0,1,0,0}
   \definecolor{YELLOW}{cmyk}{0,0,1,0}
 }

\makeatother
\begin{document}

\title{Entanglement Measures in Ion-Trap Quantum Simulators without Full Tomography}

\date{\today}

\author{J. S. Pedernales}
\affiliation{Department of Physical Chemistry, University of the Basque Country UPV/EHU, Apartado \ 644, 48080 Bilbao, Spain}
\author{R. Di Candia}
\affiliation{Department of Physical Chemistry, University of the Basque Country UPV/EHU, Apartado \ 644, 48080 Bilbao, Spain}
\author{P. Schindler}
\affiliation{Institut f\"{u}r Experimentalphysik, Universit\"{a}t Innsbruck, Technikerstra\ss e 25, A-6020 Innsbruck, Austria}
\affiliation{Department of Physics, University of California, Berkeley, CA 94720, USA}
\author{T. Monz}
\affiliation{Institut f\"{u}r Experimentalphysik, Universit\"{a}t Innsbruck, Technikerstra\ss e 25, A-6020 Innsbruck, Austria}
\author{M. Hennrich}
\affiliation{Institut f\"{u}r Experimentalphysik, Universit\"{a}t Innsbruck, Technikerstra\ss e 25, A-6020 Innsbruck, Austria}
\author{J. Casanova}
\affiliation{Department of Physical Chemistry, University of the Basque Country UPV/EHU, Apartado \ 644, 48080 Bilbao, Spain}
\author{E. Solano}
\affiliation{Department of Physical Chemistry, University of the Basque Country UPV/EHU, Apartado \ 644, 48080 Bilbao, Spain}
\affiliation{IKERBASQUE, Basque Foundation for Science, Alameda Urquijo 36, 48011 Bilbao, Spain}

\begin{abstract}
We propose a quantum algorithm in an embedding ion-trap quantum simulator for the efficient computation of $N$-qubit entanglement monotones without the necessity of full tomography. Moreover, we discuss possible realistic scenarios and study the associated decoherence mechanisms. 
\end{abstract}

\pacs{03.67.Ac, 37.10.Ty, 03.67.Mn}

\maketitle

\section{Introduction} Quantum simulators~\cite{Feynman82, Lloyd96} are currently designed with a one-to-one approach, which implies a one-to-one correspondence between the Hilbert space dimensions of the simulated system and the simulating architecture. Key examples of this approach involve the quantum simulation of black holes in Bose-Einstein condensates~\cite{Garay00}, relativistic quantum mechanical problems~\cite{Goldman09, Cirac10} and quantum phase transitions~\cite{Greiner02} in optical lattices, many-body systems with Rydberg atoms~\cite{Weimar10}, the quantum Rabi model~\cite{Ballester12} and quantum relativistic dynamics~\cite{Pedernales13} in superconducting circuits. Similar efforts have been invested in trapped-ion technologies for simulating spin models~\cite{Jane, Porras, Friedenauer08, Kim10, Lanyon11}, relativistic scattering processes~\cite{Lamata07, Gerritsma1, Casanova1, Gerritsma2, Casanova11, Lamata11}, and interacting fermionic and bosonic theories including quantum chemistry problems~\cite{ CasanovaQFT, Casanova12, Mezzacapo12, Yung14}. In this way, one-to-one quantum simulators allow us to reproduce the dynamics of a variety of quantum mechanical models. However, performing a reliable quantum simulation does not grant efficient and direct experimental access to parameters of interest as, for example, entanglement monotones~\cite{Vidal00,Horodecki09}. The usual methodology in quantum simulations reads: (a)~implementation of a certain quantum evolution, (b) infer information about the system via state tomography, and (c) compute classically relevant physical quantities out of the inferred system density matrix. Given that state tomography becomes exponentially harder with an increasing number of qubits, the one-to-one approach scales unfavorably. Thus, one may be misled to the conclusion that we are limited to scenarios where figures of merit can be directly or efficiently measured. For instance, a physical quantity such as the concurrence~\cite{Wooters98}, or its generalization to $N$ qubits~\cite{Guhne12}, is not an observable because its definition involves an antilinear operation. We have the same situation with the negativity~\cite{Vidal02}, requiring partial transpose operations. These entanglement measures need full state tomography, which becomes unfeasible for rapidly growing Hilbert-space dimensions. In this context, embedding quantum simulators (EQS) provide a mathematical framework for the efficient computation of a class of physical quantities that otherwise would require full state reconstruction~\cite{DiCandia13,Alvarez13,Pedernales14}, enhancing the capacities of one-to-one quantum simulators.

In this Letter, we provide an experimental quantum simulation recipe to efficiently compute entanglement monotones involving antilinear operations, developing the EQS concepts for an ion-trap based quantum computer. The associated quantum algorithm is composed of two steps. First, we embed the $N$-qubit quantum dynamics of interest into a larger Hilbert space involving only one additional ion qubit and stroboscopic techniques. Second, we extract the corresponding entanglement monotones with a protocol requiring only the measurement of the additional single qubit. It is noteworthy to mention that, for the computation of the associated entanglement monotones, the EQS approach does not require full-state tomography. Finally, we show how to correct experimental imperfections induced by our quantum algorithm.

Entanglement monotones are functionals of the quantum state of a system taking zero value when the state is separable, and do not increase under local operations and classical communication (LOCC). For pure states, a class of entanglement monotones can be defined as $E_\Psi(t) = | \langle \Psi | \Theta | \Psi^* \rangle | = | \langle \Psi | \Theta K | \Psi \rangle |$, where $\Theta$ is some Hermitian operator, and $K$ is the complex-conjugation operation~\cite{Osterloh1}; see, for example, the case of the two-qubit concurrence~\cite{Wooters98} where $\Theta=\sigma_y\otimes\sigma_y$. As a consequence, $E_\Psi(t)$ does not correspond to the the expectation value of a physical observable, thus it cannot be directly measured. Let us assume that we have an $N$-qubit system, represented by the wavefunction $\psi$, evolving under a Hamiltonian $H$. As described above, this system will be embedded in a larger one, requiring only one additional qubit, in such a way that $K$ becomes a physical operation~\cite{Casanova11}. The embedding procedure is based on the following mapping
\begin{equation}\label{map}
\begin{array}{ccc}
\psi&\longrightarrow& \Psi = \frac{1}{2}
\left( \begin{array}{c}
 \psi + \psi^* \\
i\psi - i\psi^*
\end{array} \right), \\ \\
H = A+iB&\longrightarrow& \tilde{H}=\big[ i \mathbb I_2 \otimes B - \sigma^y \otimes A \big].
\end{array}
\end{equation}
Here, $\psi \in \mathbb{C}^{2^N}$ and $H \in \mathbb{C}^{2^N} \times \mathbb{C}^{2^N}$ are the wavefunction and Hamiltonian (with $A$ and $B$ its real and imaginary parts)  governing the dynamics of the  $N$-qubit system in the simulated space, while $\Psi$  and $\tilde{H}$ correspond to their images in the enlarged Hilbert space, these having a dimension of $2^{N+1}$ and $2^{N+1} \times 2^{N+1}$ respectively. The matrix $M=\left(1\;,\; i\right)\otimes\mathbb{I}_{2^N}$,  projects the states of the embedding quantum simulator onto the simulated space through the identity  $\psi(t)=M \Psi(t)$. For example, for a single-qubit case where $\psi(t) = (\alpha(t), \beta(t))^T$,  $T$ being the transpose operation, the corresponding enlarged-wavefunction is  $\Psi(t) = (\alpha_{r}(t), \beta_r(t), \alpha_{i}(t), \beta_i(t))^T$, with $\alpha_{r, i}(t)$, $\beta_{r, i}(t)$ the real and imaginary parts of $\alpha(t)$ and $\beta(t)$ respectively, and the $M$ matrix is $M = \left(\begin{array}{cccc} 1& 0 & i &  0 \\ 0 & 1 & 0 & i\end{array}  \right)$. Note that due to the  property $M\tilde{H} = H M$, the previous relation is valid at any time $t$ as long as the embedding quantum simulator is initialized such that $\psi(0)=M \Psi(0)$. Although this mapping from $\psi(0)$ to $\Psi(0)$ is nonphysical, the initial state $\Psi(0)$   can be directly generated from the ground state of the simulator. In the embedding quantum simulator, the physical quantum gate  $\sigma^z \otimes \mathbb{I}_{2^N}$  applied to $\Psi$   produces a quantum state corresponding to $\psi^*$ in the simulated space, i.e. $M(\sigma_z\otimes \mathbb{I}_{2^N})\Psi(t)= \psi^*(t)$. This will allow us to efficiently compute correlations between $\psi$ and $\psi^*$ in terms of standard expectation values in the  enlarged space as follows
\begin{eqnarray}\label{anti}
\langle\psi|\Theta|\psi^*\rangle &=& \langle\Psi|M^{\dag} \Theta M (\sigma^z \otimes \mathbb{I}_{2^N})|\Psi\rangle\nonumber\\
 &=& \langle\Psi|(\sigma^z-i\sigma^x) \otimes \Theta |\Psi\rangle,
\end{eqnarray}
with $\Theta$  being a Hermitian operator. In this way, the expectation value of the antilinear operator $\Theta K$ in the simulated space  can be evaluated via the measurement of $\sigma^z \otimes \Theta$ and $\sigma^x \otimes \Theta$ in the enlarged Hilbert space. 

\section{Trapped-ion implementation} Trapped-ion systems are among the most promising technologies for quantum computation and quantum simulation protocols~\cite{Blatt08}. In such systems,  fidelities of state preparation, two-qubit gate generation, and qubit detection, exceed values of $99\%$~\cite{SchindlerNJP2013}. With current technology, more than $140$ quantum gates including many body interactions  have been performed~\cite{Lanyon11}. In this sense, the technology of trapped ions becomes a promising quantum platform to host the described embedded quantum algorithm. In the following analysis, we will rely only on a set of operations involving local rotations and global entangling M{\o}lmer-S{\o}rensen (MS) gates~\cite{Molmer99, SchindlerNJP2013}. In this sense, our method is not only applicable to trapped-ion systems. In general, it can be used in any platform where MS gates, or other long-range entangling interactions, as well as local rotations and qubit decoupling are available. Among such systems, we can mention cQED~\cite{Devoret13} where an implementation of MS gates has been recently proposed~\cite{Mezzacapo14}, or quantum photonics where MS interactions are available after a decomposition in controlled NOT gates~\cite{Obrien09}.

The embedded dynamics of an interacting-qubit system is governed by the Schr\"odinger equation $i\hbar\partial_t \Psi = \tilde{H} \Psi$, where the Hamiltonian $\tilde{H}$ is $\tilde{H}=\sum_j \tilde{H}_j$ and each $\tilde{H}_j$ operator corresponds to a tensorial product of Pauli matrices. In this way, an embedded $N$-qubit dynamics can be implemented in two steps. First, we decompose the evolution operator using standard Trotter techniques~\cite{Lloyd96,Trotter59},
\begin{equation}
\label{Trotterexpansion}
U_t = e^{-\frac{i}{\hbar}\sum_{j}\tilde{H}_jt}\approx \left(\Pi_je^{-i\tilde{H}_j t/n}\right)^n,
\end{equation}
where $n$ is the number of Trotter steps. Second, each exponential
$e^{-\frac{i}{\hbar}\tilde{H}_j t/n}$ can be implemented with a sequence of two M\o
lmer-S\o rensen gates~\cite{Molmer99}  and a
single qubit rotation between them~\cite{Mueller11, Casanova12}.  These
three quantum gates generate the evolution operator
\begin{eqnarray}\label{exp}
e^{{i \varphi \sigma^z_1 \otimes \sigma^x_2 \otimes \sigma^x_3 \ldots \otimes \sigma^x_N}},
\end{eqnarray}
where $\varphi=gt$, $g$ being the coupling constant of the single qubit rotation~\cite{Mueller11}. In Eq.~(\ref{exp}), subsequent local rotations will produce any combination of Pauli matrices.
\begin{figure}[t]
\begin{center}
\vspace{0.5cm}
\hspace{-0.3cm}
\includegraphics [width= 1.0 \columnwidth]{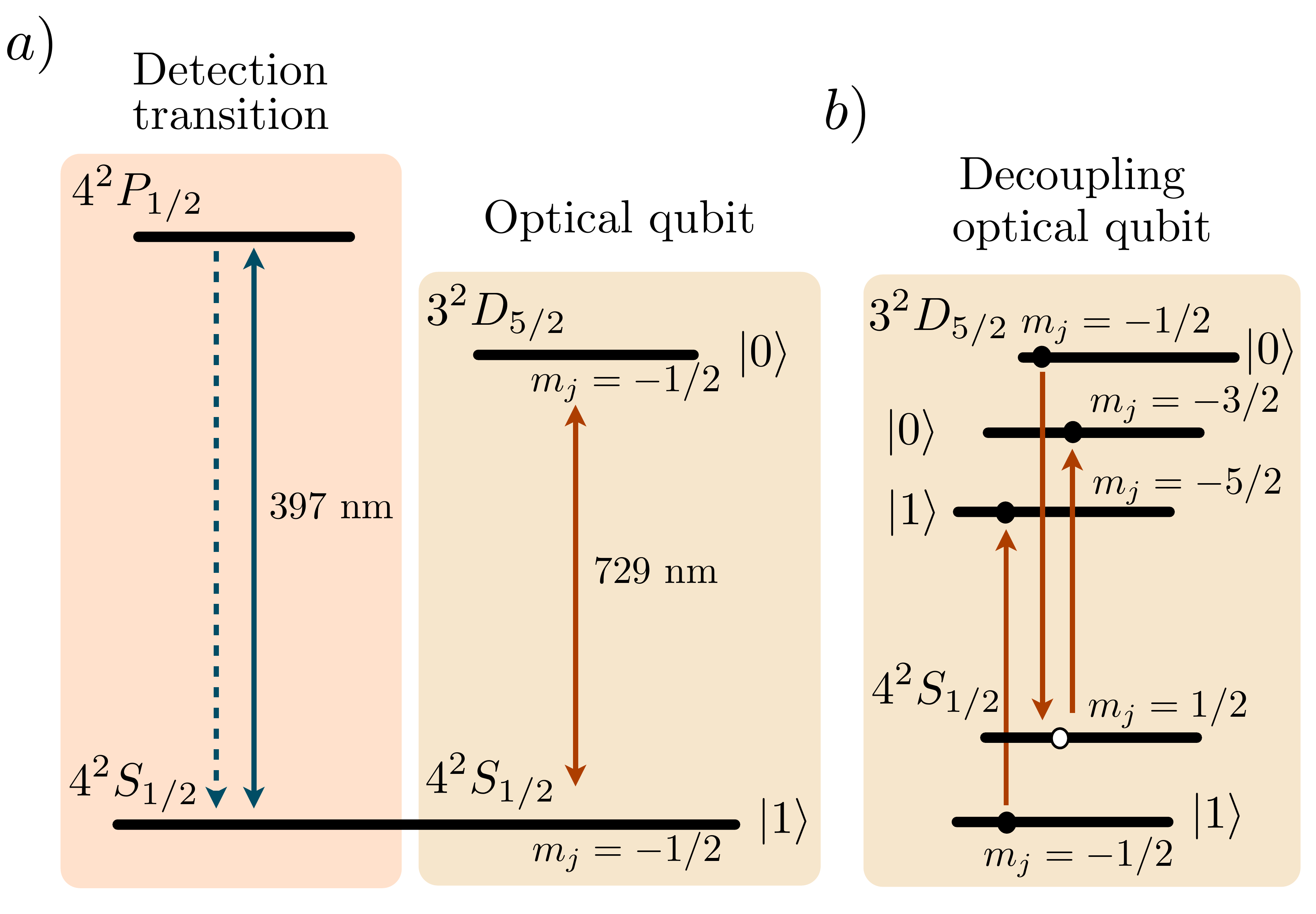}
\end{center}
\caption{(color online) a) Level scheme of $^{40}$Ca$^+$ ions. The standard optical qubit is encoded in the $m_j=-1/2$ substates of the   $3D_{5/2}$ and $4S_{1/2}$ states. The measurement is performed via fluorescence detection exciting the  $4^2 S_{1/2} \leftrightarrow 4^{2} P_{1/2}$ transition.  b) The qubit can be  spectroscopically decoupled from the evolution by shelving the information in the
 $m_j=-3/2,-5/2$ substates of the $3D_{5/2}$ state.} \label{figures levelscheme}
\end{figure}
As it is the case of quantum models involving Pauli operators, there exist different representations of the same dynamics. For example, the physically equivalent Ising Hamiltonians, $H_1=\omega_1 \sigma^x_1  + \omega_2  \sigma^x_2 + g \sigma^y_1 \otimes \sigma^y_2$ and $H_2 = \omega_1 \sigma^y_1 + \omega_2  \sigma^y_2 + g \sigma^x_1 \otimes \sigma^x_2$, are mapped onto the enlarged space as $\tilde{H}_1=-\omega_1 \sigma^y_0 \otimes \sigma^x_1 - \omega_2 \sigma^y_0 \otimes \sigma^x_2 - g \sigma^y_0 \otimes \sigma^y_1 \otimes \sigma^y_2$ and  $\tilde{H}_2= \omega_1  \sigma^y_1 + \omega_2 \sigma^y_2 -  g \sigma^y_0 \otimes \sigma^x_1 \otimes \sigma^x_2$. In principle, both Hamiltonians $\tilde{H}_1$ and $\tilde{H}_2$ can be implemented in trapped ions. However, while $\tilde{H}_1$ requires two- and three-body interactions, $\tilde{H}_2$ is implementable with a collective rotation applied to the ions $1$ and $2$ for the implementation of the free-energy terms, and MS gates for the interaction term. In this sense, $\tilde H_2$ requires less experimental resources for the implementation of the EQS dynamics. Therefore, a suitable choice of the system representation can considerably enhance  the performance of the simulator.

\section{Measurement protocol}  We want to measure correlations of the form appearing in Eq.~(\ref{anti}), with $\Theta$ a linear combination of tensorial products of Pauli matrices and identity operators. This information can be encoded in the expectation value $\langle \sigma_a^\alpha \rangle$ of one of the ions in the chain after performing two evolutions of the form of Eq.~(\ref{exp}).  Let us consider the  operators $U_1=e^{-i\varphi_1(\sigma^i_1 \otimes \sigma^j_2 \otimes \sigma^k_3 ...)}$ and $U_2=e^{-i\varphi_2(\sigma^o_1\otimes \sigma^p_2 \otimes \sigma^q_3 ... )}$ and choose the Pauli matrices $\sigma^i_1, \sigma^j_2, ...$ and $ \sigma^o_1, \sigma^p_2, ... $ such that $U_1$ and $U_2$ commute and both anticommute with the Pauli operator to be measured $\sigma^\alpha_a$. In this manner, we have that
\begin{eqnarray}
\langle \sigma^\alpha_a \rangle_{\varphi_1, \varphi_2 =\frac{\pi}{4}} &=& \langle U_1^{\dag}(\frac{\pi}{4})U_2^\dag(\frac{\pi}{4}) \ \sigma^\alpha_a \  U_1(\frac{\pi}{4}) U_2(\frac{\pi}{4}) \rangle\nonumber\\
&=& \langle \sigma^i_1 \sigma^o_1 \otimes \sigma^j_2\sigma^p_2  \otimes ... \otimes \sigma_a^\alpha \sigma_a^l \sigma_a^r... \rangle.
\end{eqnarray}
Then, a suitable choice of Pauli matrices will produce the desired correlation. Note that this protocol always results in a correlation of an odd number of Pauli matrices. In order to access a correlation of an even number of qubits, we have to measure a two-qubit correlation $\sigma^\alpha_a \otimes \sigma^\beta_b$ instead of just $\sigma_a^\alpha$. For the particular case of correlations of only Pauli matrices and no identity operators, evolution $U_2$ is not needed and no distinction between odd and even correlations has to be done. For instance, if one is interested in an even correlation like ${\sigma^y_1 \otimes \sigma^x_2 \otimes \sigma^x_3 \otimes \sigma^x_4 \otimes \mathbb{I}_5 \otimes ... \otimes \mathbb{I}_N}$, $N$ being the number of ions of the system, then one would have to measure observable
$\sigma_1^y \otimes \sigma_2^x$ after the  evolutions  $U_1=e^{-i(\sigma_1^x \otimes \sigma_2^y \otimes \sigma_3^y \otimes \sigma_4^y \otimes \sigma_5^y \otimes ... \otimes \sigma_j^y \otimes ...) \varphi}$ and ${U_2=e^{-i(\sigma_1^x \otimes \sigma_2^y \otimes \sigma_3^z \otimes \sigma_4^z \otimes \sigma_5^y \otimes ...\otimes \sigma_j^y \otimes ...)\varphi}}$. However, for the particular case of $N=4$ a single evolution $U_1=e^{-i(\sigma_1^x \otimes \sigma^x_2 \otimes \sigma^x_3
\otimes \sigma^x_4) \varphi}$ and subsequent measurement of $\langle \sigma_1^z \rangle$ is enough. Note that all the gates in the protocol, as they are of the type of Eq.~(\ref{exp}), are implementable with single qubit and MS gates.

\section{Examples} Consider the Ising Hamiltonian for two spins, ${ H=\hbar\omega_1 \sigma^y_1 + \omega_2 \sigma^y_2 + g \sigma^x_1 \otimes \sigma^x_2 }$ whose image in the enlarged space corresponds to $ \tilde H = \omega_1 \sigma^y_1 + \omega_2 \sigma^y_2 - g \sigma^y_0 \otimes \sigma^x_1 \otimes \sigma^x_1 $. The evolution operator associated to this Hamiltonian can be implemented using the Trotter method from Eq.~(\ref{Trotterexpansion}) with $(\tilde{H}_{1}, \tilde{H}_2, \tilde{H}_3) =(\omega_1 \sigma^y_1, \omega_2 \sigma^y_2, - g \sigma^y_0 \otimes \sigma^x_1 \otimes \sigma^x_2)$. While evolutions $e^{-\frac{i}{\hbar}\tilde{H}_1 t/n}$ and $e^{-\frac{i}{\hbar}\tilde{H}_2 t/n}$ can be implemented with single ion rotations, the evolution $e^{-\frac{i}{\hbar}\tilde{H}_3 t/n}$, which is of the kind described in Eq.~(\ref{exp}), is implemented with two MS gates and a single ion rotation. This simple case allows us to compute directly quantities such as the concurrence measuring $\langle \sigma_0^z \sigma_1^y \sigma_2^y \rangle$ and $\langle \sigma_0^x \sigma_1^y \sigma_2^y \rangle$. According to the measurement method introduced above, to access these correlations we first evolve the system under the gate $U=e^{-i(\sigma_0^y \otimes \sigma_1^y \otimes \sigma_2^y)\varphi}$ for a time such that $\varphi=\frac{\pi}{2}$, and then measure $\langle \sigma_0^x \rangle$ for the first correlation and $\langle \sigma_0^z \rangle$ for the second one.

\begin{figure}[t]
\begin{center}
\vspace{0.5cm}
\hspace{-0.5cm}
\includegraphics [width= 1.05 \columnwidth]{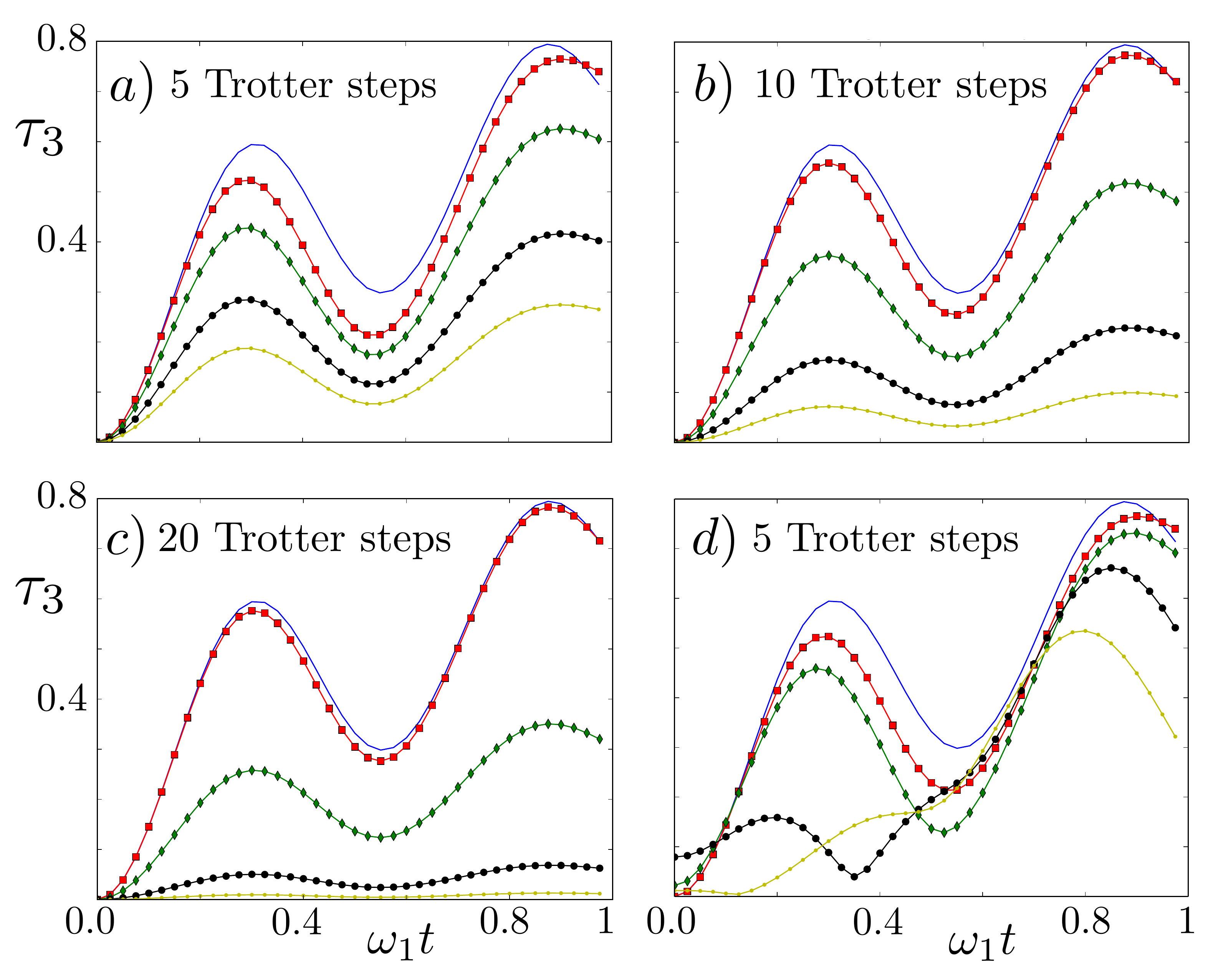}
\end{center}
\caption{(color online) Numerical simulation of the $3$-tangle evolving under Hamiltonian in Eq.~($\ref{EnlargedIsing3qubit}$) and assuming different error sources. In all the plots the blue line shows the ideal evolution. In a), b), c) depolarizing noise is considered, with N=5,10 and 20 Trotter steps, respectively. Gate fidelities are $\epsilon =1, 0.99$, $0.97$, and $0.95$ marked by red rectangles, green diamonds, black circles and yellow dots, respectively. In d) crosstalk between ions is added with strength $\Delta_0 = 0, 0.01$, $0.03$, and $0.05$ marked by red rectangles, green diamonds, black circles and yellow dots, respectively. All the simulations in d) were performed with 5 Trotter steps. In all the plots we have used $\omega_1=\omega_2 = \omega_3=g/2=1$.} \label{figures simulation}
\end{figure}

Based on the two-qubit example, one can think of implementing a three-qubit model as $ H_{\rm GHZ}= \omega_1 \sigma^y_1 + \omega_2 \sigma^y_2 + \omega_3 \sigma^y_3 + g \sigma^x_1 \otimes \sigma^x_2 \otimes \sigma^x_3 $, which in the enlarged space corresponds to
\begin{equation}
 \label{EnlargedIsing3qubit}
 \tilde H_{\rm GHZ} = \omega_1 \sigma^y_1 +  \omega_2 \sigma^y_2 + \omega_3 \sigma^y_3 - g \sigma^y_0 \otimes \sigma^x_1 \otimes \sigma^x_2 \otimes \sigma^x_3.
\end{equation}
This evolution results in GHZ kind states, which can be
readily detected using the 3-tangle $\tau_3$~\cite{Dur00}. This is an entanglement monotone of the general class of  Eq.~(\ref{anti}) that can be computed in the enlarged space by measuring  $\big|-\langle\tilde\psi(t)|\sigma_z\otimes\mathbb{I}_2\otimes\sigma_y\otimes\sigma_y-i\sigma_x\otimes\mathbb{I}_2\otimes\sigma_y\otimes\sigma_y|\tilde\psi(t)\rangle^2 +\langle\tilde\psi(t)|\sigma_z\otimes\sigma_x\otimes\sigma_y\otimes\sigma_y-i\sigma_x\otimes\sigma_x\otimes\sigma_y\otimes\sigma_y|\tilde\psi(t)\rangle^2  + \langle \tilde \psi(t) | \sigma_z \otimes \sigma_z \otimes \sigma_y \otimes \sigma_y - i \sigma_x \otimes \sigma_z \otimes \sigma_y \otimes \sigma_y | \tilde \psi(t) \rangle^2 \big| $.
More complex Hamiltonians with interactions involving only three of the four particles can also be implemented.  In this case, the required entangling operations acting only on a part of the entire register can be realized with the aid of splitting the MS operations into smaller parts and
inserting refocusing pulses between them as shown in Ref.~\cite{Mueller11}. An alternative method is
to decouple the spectator ions from the laser light by shelving the quantum information into additional Zeeman substates of the ions as sketched in Fig.~(\ref{figures levelscheme}) for $^{40}$Ca$^+$ ions.
This procedure has been successfully demonstrated in Ref.~\cite{SchindlerNatPhys13}.
For systems composed of a larger number of qubits, for example $N> 10$, our method yields nontrivial results given that the standard computation of  entanglement monotones of the kind  $\langle \psi(t)| \Theta | \psi(t)^*\rangle$ requires the measurement of a number of observables that grows exponentially with $N$. For example, in the case of $\Theta = \sigma^y \otimes \sigma^y \otimes \dots \otimes \sigma^y$~\cite{Osterloh1} our method requires the evaluation of $2$ observables while the standard procedure based on state tomography requires, in general, the measurement of $2^{2N}-1$ observables.

\section{Experimental considerations} A crucial issue of a quantum simulation algorithm is its susceptibility to experimental imperfections. In order to investigate the deviations with respect to the ideal case, the system dynamics needs to be described by completely positive maps instead of unitary dynamics. Such a map is defined by the process matrix $\chi$ acting on a density operator $\rho$ as follows: $\rho \rightarrow \sum_{i,j} \chi_{i,j} \sigma^i \rho \sigma^j,$ where $\sigma^i$ are the Pauli matrices spanning a basis of the operator space.  In complex algorithms, errors can be modeled by adding a depolarizing process  with a probability
$1 - \varepsilon$ to the ideal process $\chi^{id}$
\begin{equation}
\label{depnoisemodel}
\rho \rightarrow \varepsilon \sum_{i,j} \chi^{id}_{i,j} \sigma^i \rho \sigma^j + (1- \varepsilon) \frac{\mathbb{I}}{2^N}  .
\end{equation}
In order to perform a numerical simulation including this error model, it is required to decompose the quantum simulation into an implementable gate sequence. Numerical simulations of the Hamiltonian in Eq.~(\ref{EnlargedIsing3qubit}), including realistic values gate fidelity $\varepsilon = \{1, 0.99, 0.97, 0.95\}$ and for $\{5,10,20\}$ Trotter steps, are shown in Figs.~\ref{figures simulation} a), b), and c).
Naturally, this analysis is only valid if the noise in the real system is close to depolarizing noise. However, recent analysis of entangling operations indicates that this noise model is accurate~\cite{Ozeri13, SchindlerNJP2013}.  According to Eq. (\ref{depnoisemodel}), after $n$ gate operations, we show that
\begin{equation}
\label{idealexpectation}
\langle O \rangle_{\mathcal{E}_{id}(\rho)}=\frac{\langle O\rangle_{\mathcal{E}(\rho)}}{\varepsilon^n}-\frac{1-\varepsilon^n}{\varepsilon^n}\text{Tr}\,(O),
\end{equation}
where $\langle O \rangle_{\mathcal{E}_{id}(\rho)}$ corresponds to the ideal expectation value in the absence of decoherence, and $\langle O\rangle_{\mathcal{E}(\rho)}$ is the observable measured in the
experiment. Given that we are working with observables composed of tensorial products of Pauli operators $\sigma^y_0\otimes\sigma_1^x ...$ with $\text{Tr}\,(O)=0$, Eq.~(\ref{idealexpectation}) will simplify to $\langle O \rangle_{\mathcal{E}_{id}(\rho)}=\frac{\langle O\rangle_{\mathcal{E}(\rho)}}{\varepsilon^n}$. In order to retrieve with uncertainty $k$ the expected value of an operator $O$, the experiment will need to be repeated $N_{emb}= \left( \frac{1}{k \epsilon^n} \right)^2$ times. Here, we have used $k\equiv\sigma^{\mathcal{E}(\rho)}_{\langle O \rangle}=\sigma^{\mathcal{E}(\rho)}_O/\sqrt{N}$ ( for large $N$), and that the relation between the standard deviations of the ideal and experimental expectation values is $\sigma^{\mathcal{E}(\rho)}_O=\sigma^{\mathcal{E}_{id}(\rho)}_O/\varepsilon^n$. If
we compare $N_{emb}$ with the required number of repetitions to measure the same entanglement monotone to the same accuracy $k$ in a one-to-one quantum simulator, $N_{oto} = 3^{N_{qubits}} \left(\frac{1}{k \delta^{n}}\right)^2$, we have
\begin{equation}
\label{tomoVSemb}
\frac{N_{emb}}{N_{oto}}=l\left(\frac{\delta}{ \sqrt{3}\varepsilon}\right)^{2N_{qubit}}.
\end{equation}
Here, $l$ is the number of observables corresponding to a given entanglement monotone in the enlarged space, and $\delta$ is the gate fidelity in the one-to-one approach. We are also asuming that full state tomography of $N_{qubits}$ qubits requires $3^{N_{qubit}}$ measurement settings for experiments exploiting single-qubit discrimination during the measurement process~\cite{Roos04}. Additionally, we assume the one-to-one quantum simulator to work under the same error model but with $\delta$
fidelity per gate. Finally, we expect that the number of gates grows linearly with the number of qubits, that is $n\sim N_{qubit} $, which is a fair assumption for a nearest-neighbor interaction model. In general, we can assume that $\delta$ is always bigger than $\varepsilon$ as the embedding quantum simulator requires an additional qubit which naturally could increase the gate error rate. However, for realistic values of $\varepsilon$ and $\delta$, e.g. $\varepsilon=0.97$ and $\delta=0.98$ one can prove that $\frac{N_{emb}}{N_{oto}}\ll 1$. This condition is always fulfilled for large systems if $\frac{\delta}{ \sqrt{3}\varepsilon}<1$. The latter is a reasonable assumption given that in any quantum platform it is expected $\delta\approx\epsilon$ when the number of qubits grows, i.e. we expect the same gate fidelity for $N$ and $N+1$ qubit systems when $N$ is large.  Note that this analysis assumes that the same amount of Trotter steps is required for the embedded and the one-to-one simulator. This is a realistic assumption if one considers the relation between $H$ and $\tilde{H}$ in Eq.~(\ref{map}). A second type of imperfections are undesired unitary operations due to imperfect calibration of the applied gates or due to crosstalk between neighboring qubits. This crosstalk occurs when performing operations on a single ion due to imperfect single site illumination~\cite{SchindlerNJP2013}. Thus the operation $s_j^{z}(\theta)=\exp(-i \, \theta \, \sigma_{j}^{z}/ 2 )$ needs to be written as $s_j^{z}(\theta)=\exp(-i \, \sum_k \epsilon_{k,j} \, \theta \sigma_k^z / 2 )$ where the crosstalk is characterized by the matrix $\Delta$. For this analysis, we assume that the crosstalk affects only the nearest neighbors with strength $\Delta_0$ leading to a matrix $\Delta=\delta_{k,j} + \Delta_0 \, \delta_{k\pm1,j} $. In Fig.~\ref{figures simulation} d) simulations including crosstalk are shown. It can be seen, that simulations with increasing crosstalk show qualitatively different behavior of the 3-tangle, as in the simulation for $\Delta_0=0.05$ (yellow line) where the entire dynamics is distorted. This effect was not observed in the simulations including depolarizing noise and, therefore, we identify unitary crosstalk as a
critical error in the embedding quantum simulator. It should be noted that, if accurately characterized, the described crosstalk can be completely compensated experimentally~\cite{SchindlerNJP2013}.

In conclusion, we have proposed an embedded quantum algorithm for trapped-ion systems to efficiently compute entanglement monotones for $N$ interacting qubits at any time of their evolution and without the need for full state tomography. It is noteworthy to mention that the performance of EQS would outperform similar efforts with one-to-one quantum simulators, where the case of 10 qubit may be considered already as intractable. Furthermore, we showed that the involved decoherence effects can be corrected if they are well characterized. We believe that EQS methods will prove useful as long as the Hilbert-space dimensions of quantum simulators grows in complexity in different quantum platforms.

The authors acknowledge support from Spanish MINECO FIS2012-36673-C03-02; UPV/EHU UFI 11/55; UPV/EHU PhD grant; Basque Government IT472-10; the Austrian Science Fund (FWF), 
through the SFB FoQus (FWF Project No. F4002-N16), as well as the Institut f\"ur Quanteninformation GmbH; and CCQED, PROMISCE, SCALEQIT, and QUASIRIO EU projects. This research was funded by the Office of the Director of National Intelligence (ODNI), Intelligence Advanced Research Projects Activity (IARPA), through the Army Research Office grant W911NF-10-1-0284. All statements of fact, opinion or conclusions contained herein are those of the authors and should not be construed as representing the official views or policies of IARPA, the ODNI, or the U.S. Government.

\end{document}